\documentclass{llncs}

\usepackage{amsmath}
\usepackage{amssymb}
\usepackage{array}
\usepackage{listings}
\usepackage{fancyvrb}
\usepackage{graphicx}
\usepackage{enumerate}
\usepackage{breakcites}

\newcommand{\cscparagraph}[1]{\paragraph*{\normalfont \textbf{#1}}}

\newcommand*\R{\mathrel\mathcal{R}}
\newcommand*\Rp{\mathrel\mathcal{R}^{\prime}}
\newcommand*\PMathML{Presentation MathML}
\newcommand*\CMathML{Content MathML}

\title{A Survey on Retrieval of Mathematical Knowledge%
\thanks{The final publication is available at http://link.springer.com.}}

\author{Ferruccio Guidi, Claudio Sacerdoti Coen}

\institute{Department of Computer Science and Engineering -- DISI \\
           University of Bologna\\
           \email{\{ferruccio.guidi,claudio.sacerdoticoen\}@unibo.it}}

\begin{document}
\maketitle

\begin{abstract}
We present a short survey of the literature on indexing and retrieval
of mathematical knowledge, with pointers to 72 papers and tentative taxonomies of both retrieval problems and
recurring techniques.
\end{abstract}

\section{Purpose Driven Taxonomy of Retrieval Problems}
Retrieval of mathematical knowledge is always presented as the
low hanging fruit of Mathematical Knowledge Management, and it has been addressed
in several papers by people coming either from the formal methods or from the information retrieval
community.
The problem being resistant to classical content search techniques \cite{LarsonBerkeley}, it is usually addressed combining a small set of new
ideas and techniques that are recurrent in the literature.
Despite the amount of work, however, there is not a single solution that is the clearly winning on the others,
nor convincing unbiased benchmarks to compare solutions.
Some authors like \cite{DBLP:conf/mkm/KohlhaseK07} also suggest that the community should first better understand
the actual needs of mathematicians from an unbiased perspective to improve the MKM technology as a whole.
In this paper we collect a hopefully comprehensive bibliography,
and we roughly classify the papers according to novel taxonomies both for
the problems and the techniques employed.
The only other surveys on the same topic are \cite{Asperti04searchingmathematics}, now outdated and focused mostly on (European) research projects that contributed to the topic in the 6th Framework Programme, \cite{zanibbi2012recognition}, which covers less literature in much greater detail without attempting a classification, \cite{1083259}, which is focused on evaluation of mathematics retrieval, and \cite{bachliska}, which is written in Slovak.

We begin our discussion with a purpose driven taxonomy made of three different retrieval problems that deal with mathematical knowledge. Each problem is characterised by its own set of expectations and constraints,
and adopting a solution to another problem may be infeasible or yield poor results. In the next sections we classify the papers according to an encoding based taxonomy (presentation vs content vs semantics) and to a taxonomy of techniques employed. Finally we point to the rich literature relative to the problem of ranking, and we touch the problem of evaluation of systems. We conclude with some notes on the availability of math retrieval systems.

\clearpage
\cscparagraph{Problem 1: Document Retrieval}~\\

\textbf{Objective:}
A \emph{human} is interested in recalling a \emph{set of mathematical documents} (or fragments)
that are related to a particular mathematical topic.
Typically it is not the case that only one document provides the correct answer;
on the contrary the user may be interested in a corpora of different documents that yield different,
only partially overlapping information.
In \cite{DBLP:conf/mkm/Kohlhase14} and other papers there are attempts at a classification of the information needs
of users. However, at the moment only the system described in \cite{zhao2008math} tries to use the classification
to improve the user experience.

\textbf{Input:}
The human composes a query combining keywords (e.g. for topics \cite{adeel2008math}),
free text and mathematical formulae.
Often the mathematical formulae are intended as examples of expressions related to the topic of interest.
For example, a user interested in trigonometric identities can just enter one identity to retrieve them all.
Or a formula showing a particular property of a special function can be used to disambiguate the special function
among the ones with similar names.

The query can be composed using a very simple, Google-inspired, single line interface,
or written using an ad-hoc query language
(see \cite{DBLP:conf/caine/AltamimiY07,DBLP:journals/mics/AltamimiY08,DBLP:conf/sede/YoussefA07}
for some proposals), or by filling in some form.
The first solution is the one preferred in the literature.
In \cite{DBLP:conf/mkm/Kohlhase14}
a comparison of the behaviour of mathematicians vs other users highlighted that
the professional mathematician is more interested in the precision of the output than the effort put into the input.
Therefore mathematicians may use and appreciate more complex interfaces.
On the contrary, other users are likely to prefer a simple, modern search interface.

Formulae can be entered in some textual syntax (e.g. \LaTeX{}, MathML), maybe with the help of on-the-fly formulae
rendering \cite{DBLP:conf/mkm/LiskaSR14}, or using graphical editors \cite{munavalli2006mathfind},
or they can be acquired from hand-written snippets \cite{zbMATH05607874}.
The formula is likely to contain errors and ambiguities,
for example if it is encoded at the presentational level (e.g. in \PMathML{} or \LaTeX{}),
if it is acquired from hand-written text, or if the user only remembers it partially or in a wrong way.
Errors and ambiguities are not a critical problem because
formulae are just used to retrieve documents that contain \emph{similar} formulae according to some similarity
criterion. In \cite{zbMATH05607874} the authors address the problem of combining and ranking results from different
queries generated from ambiguous formulae due to errors in the recognition process.
See also \cite{zanibbi2012recognition} for a survey on the interaction between mathematical information retrieval
and mathematical document recognition.
Some authors \cite{zbMATH05621232}
suggest that the visual presentation of the formula may sometimes be important in the definition of similarity,
whereas in other situations it is the mathematical \emph{content} of the formula that matters.
Logically equivalent formulae whose content encoding is highly
different are better considered less similar.

Once the search engine returns the result,
the user may be given the opportunity to enhance the query by further filtering.

\textbf{Output:}
the output is a \emph{ranked list} of matching documents or document fragments
(e.g. a chapter of a book, or a section of an article).
When the query involves mathematical formulae, the ranking is determined by the similarity relation.
The user must be given the possibility to quickly determine whether the matched document is interesting or not.
Therefore the problem of how to present summaries of the selected documents in the result list
is of fundamental
importance \cite{LískaSojkaRuzickaMravec11,DBLP:conf/mkm/LiskaSR14,zbMATH05607875,WolskaGrigore10,DBLP:conf/iasse/Youssef05,DBLP:conf/mkm/Youssef06,DBLP:conf/mkm/Youssef07,DBLP:journals/mics/Youssef08}.
Even highlighting correctly the bits of the summary that matches the query
can make a significant difference in the user
experience \cite{LískaSojkaRuzickaMravec11,DBLP:conf/mkm/LiskaSR14,DBLP:conf/iasse/Youssef05,DBLP:conf/mkm/Youssef06}.
The list of results must be the starting point for further investigations by the user.
At least, all results must contain hyperlinks or other ways to retrieve the original document the summary points to.
A study of user requirements in \cite{zhao2008math} suggests that results should be presented after clustering them
according to their resource type (research paper, tutorial, slides, course, book, etc.).
For example, a student may immediately decide to skip research papers,
and a researcher may skip websites and tutorials.

\textbf{Constraints:}
a balance must be obtained between \emph{precision} (the fraction of retrieved documents that are relevant)
and \emph{recall} (the fraction of relevant documents that are retrieved).
To maximise recall, precision is affected and many out of topic documents (false positives) are retrieved,
penalising performance.
Too many results are overwhelming and the user is likely to give attention only to the first ones in the list.
Therefore, the search engine does not have to rank and produce summaries of documents with low scores.
The ranking function is ultimately the one responsible for the perceived quality of the search engine.

Since the query is intended to be issued by a human,
the performance of the search engine is not a critical requirement and up to a few seconds
(or even minutes in some particular situations) may be acceptable.
Nevertheless, modern textual search engines like Google are extremely fast,
and the user is likely to expect the queries to be solved in less than a second.

\cscparagraph{Problem 2: Formula Retrieval}~\\

\textbf{Objective:}
A program --- more rarely a human --- is interested in retrieving \emph{all} formulae
that are in some relation $\R$ with a query formula $E$.
Sometimes the formula $E$ can actually be a set of formulae.
For example, $E$ can be a goal to be proved automatically,
and $T \R E$ when $T$ is the conclusion of the statement of a theorem
that can be instantiated to prove $E$.
More precisely, $T$ contains metavariables to be instantiated and $\R$ is one-sided unification
up to some equational theory.
The dual query is also used in the literature:
$E$ is a property (a statement containing metavariables),
$\R$ is unification and the query finds all operations that satisfy the property $E$.
By using several properties at once, the query can find all models of a given theory
(e.g. all semirings in the library) \cite{DBLP:conf/mkm/NormannK07},
also up to renaming of constants and properties.
An interesting application presented in \cite{DBLP:conf/mkm/GauthierK14},
that uses techniques similar to \cite{DBLP:conf/mkm/NormannK07}, consists in matching concepts across libraries
by first computing properties of an object in one library (i.e. patterns like commutativity of a binary operator)
and then looking for objects in the other library that satisfy the same properties.
To be more effective, properties are extracted from all libraries and concepts are matched according to a similarity
measure to identify objects that satisfy a similar set of properties.
A third example is obtained by choosing logical implication for $\R$.
The query looks for all formulae that imply $E$.

\textbf{Input:}
one or more formulae $E$ that may or may not contain metavariables to be instantiated.
Rarely, additional constraints can be expressed using keywords, classifications, free text, authors, etc.
Formulae are not supposed to be ambiguous or contain errors.
In \cite{asperti2006content} ambiguity is resolved before performing the query using type checking and interaction
with the user.

Some dedicated query languages are proposed to specify the structure of the formulae
$E$~
\cite{bancerek2006information,DBLP:conf/mkm/BancerekR03,DBLP:conf/mkm/BancerekU04,%
DBLP:conf/mkm/GuidiS03,Kamali:2010:NMR:1871437.1871635,DBLP:conf/aisc/Rabe12}.
They are implemented on top of relational databases or ad-hoc in-memory indexes.

\textbf{Output:}
the query is meant to retrieve a set of formulae that satisfy a certain property.
When the search is performed by a program,
there is no need to present summaries of the document the formula occurs in.
Even when a human issued the search, an hyperlink to the document may be sufficient.

In many situations the relation $\R$ can be extended to a ternary relation $T \R_\rho E$
meaning that $T$ is related to $E$ with score $\rho$, and the results can be ranked according to $\rho$.
For example, if $\R$ reduces a proof of $E$ to a proof of $T$,
then the $T$'s may receive a higher score if they are judged easier to prove.

\textbf{Constraints:}
maximisation of the recall is fundamental.
The query should return all formulae that satisfy the query, even if they rank very low.
Because searches are often basic operations of complex algorithms (e.g. automatic provers), speed is also critical.
In several situations, the searches need to be performed in milliseconds.

To speed up the searches or when the relation $\R$ is undecidable,
the search engine may use a second decidable relation $\Rp$ such that $\R \subseteq \Rp$.
Using $\Rp$, the query can return false positives,
i.e. formulae $T$ such that $T \Rp E$ but not $T \R E$.
For example, when $E$ is a pattern and $\R$ is unification,
$\Rp$ may ignore the structure of the two formulae $E$ and $T$ and conclude $E \Rp T$
when all symbols in $E$ are also in $T$.
Example: $f(xz,y+z) \R f(?,?+?)$ and $f(y+z,xz) \Rp f(?,?+?)$,
but not $f(y+z,xz) \R f(?,?+?)$.

\cscparagraph{Problem 3: Document Synthesis}~\\

\textbf{Objective:}
Composing a new mathematical document assembling fragments to be retrieved from a library.
The most common occurrence is in educational software
where a learning object must be assembled according to the expertise of the user, the topic of interest,
etc. \cite{baumgartner2003automated,DBLP:conf/aisc/LibbrechtDMLDH08,libbrecht2006methods}.
An unrelated example is the automatic generation of summaries and statistics for a mathematical
library \cite{DBLP:conf/mkm/BancerekR03}.
A final example is mining of formalised libraries,
for example to build visual representations of the graph of dependencies over an axiom
(to understand its implications) or a definition/statement (to understand the propagation of changes).

A variant is to solve a mathematical problem by composing mathematical (Web)
services \cite{DBLP:conf/mkm/CaprottiDT04}.
Each service exposes metadata about the problem solved, the algorithm
implemented, and its preconditions and postconditions.

\textbf{Input:}
the query is not likely to involve mathematical formulae,
and it is usually expressed using a query language over ontologies.
A high level interface may hide the underlying query language.
Sometimes the query is fixed once and for all, and needs to be run at regular periods.

\textbf{Output:}
the expected output depends on the particular use case and
it is usually made of a single result in place of a ranked list.
The result may consist of a graph of objects and relations between them,
or it may consist of the minimal information to build the expected document or solve the algorithmic problem.

\textbf{Constraints:}
the constraints depend on the particular use case.

~\vspace{-0.6cm}\\

\noindent
After an initial screening of the literature, we decided to analyse only papers about the first two problems,
where formulae play a central role.
Indeed, at a first glance most solutions to Document Synthesis employ standard query languages for ontologies,
and only the ontologies themselves are math-specific
(e.g. \cite{DBLP:conf/mkm/CaprottiDT04,DBLP:conf/aisc/LibbrechtDMLDH08}).
Logic programming languages are also employed to represent what the user knows/ignores and the inference rules to
assemble documents \cite{baumgartner2003automated}.

Moreover, we did not find in the literature convincing examples for
the need of very expressive query languages to solve the Document Retrieval and the Formula Retrieval problems, where the kinds of queries are essentially
fixed a priori.
Moreover, evaluation of queries expressed in these languages are reported to be too slow to be used for Formula Retrieval.
Sometimes additional techniques are employed for Document Synthesis,
like semantical query reduction to relax the user provided query by allowing additional topics
close to the one specified by the user \cite{DBLP:conf/mkm/Libbrecht13}.
These techniques too seem to be very general and applicable to domains very different from that of mathematics.

Despite the strong interest of the community in the use of formulae in
queries, studies on the behaviour of users \cite{DBLP:conf/mkm/KohlhaseK07,zhao2008math} conclude that
the added value may be low, and the finding is confirmed
in \cite{libbrecht2006methods,DBLP:conf/mkm/Miller13}
where the logs of the DLMF and of ActiveMath search engines are analysed concluding that only few queries contain
mathematical formulae, most are very simple ones, and such queries do not yield satisfactory results.

\section{Encoding Based Taxonomy}\label{sec:mathml}
Mathematical information can be encoded in a library at three different levels.
The most shallow one is \emph{presentation}.
Presentation markup uses a finitary language to express the bi-dimensional layout of a formula,
useful to present it to the reader.
The standard XML language for presentation is \emph{\PMathML},
and several tools can generate \PMathML{} from \LaTeX (e.g. LaTeXML, Tralics), PDF files (e.g. Maxtract),
handwritten text (e.g. InfTy Reader),
digitised documents (e.g. InfTy Reader) or content markup (e.g. via XSLT stylesheets).
We can therefore assume that the totality of the documents to be indexed are available in \PMathML{}.

The next level is \emph{content}.
At the content level, the structure of the formula is described,
and symbols and operators appearing in it are linked to their entry in an ontology,
called \emph{content dictionary} in OpenMath terminology.
The markup language is finitary, but the ontology is not since new mathematical entities can always be defined.
The relation between content and presentation is one-to-many:
the same presentation markup may represent different content expressions (\emph{ambiguity}),
and a content expression can be given different presentations
according to the conventions of the community of readers, the language of the reader,
but also for purely aesthetic reasons like constraints on the size of the formula.
OpenMath and \CMathML{} are the two standard XML language for formulae at the content level.
OMDoc is an attempt at standardising at the content level whole mathematical documents, comprising proofs.
Content markup is currently mostly used for the exchange of formulae between systems, in particular CAS.
There are no significant examples of large libraries of documents natively written using content markup.
Nevertheless, there are tools like SnuggleTeX 
based on heuristics to semantically enrich
(annotated) \LaTeX{} documents or even MathML Presentation documents to content.

The last level is \emph{semantics} and it is specific to libraries of formalised mathematical knowledge.
The semantics level refines the content level by picking for every content level object
one particular definition in a given logic. The definition chosen embeds the object with additional properties,
e.g. computational properties.
For example, addition over natural numbers can be defined in the Calculus of Inductive Constructions
as a non-computable ternary predicate in logic programming style, or as a recursive function on the first argument
--- such that $0+x$ and $x$ become logically indistinguishable ---
or as a recursive function on the second argument ---
so that $0+x$ and $x$ are not indistinguishable, but only provably equal.
Interactive theorem provers often provides an XML dump of their internal semantics representation.

Formula Retrieval is always formulated either on semantics markup or on content markup.
Even when the semantic markup is available,
it may be convenient to convert the library to content level
by identifying alternative definitions of the same mathematical notion.
In this way, it becomes possible to retrieve useful theorems on
mathematically equivalent definitions,
in the hope to reuse them after conversion to the definitions in use.
One application of this technique is reuse of libraries across different systems based on the same logic.

The Document Retrieval problem is formulated in a way that is agnostic of the encoding.
However, the user is likely to enter formulae in the query using a
presentation language (mostly \LaTeX{}, even if MathML starts to be
used \cite{LískaSojkaRuzickaMravec11,DBLP:conf/mkm/LiskaSR14,zbMATH05607875}).
Some authors have provided evidence that precision is improved when exploiting parallel markup,
even when the content part is automatically generated from the presentation
part \cite{DBLP:conf/mkm/NghiemKTA14}.
See \cite{DBLP:conf/aisc/MillerY08} for motivations against content/parallel markup and
in favour of a more lightweight encoding of content information in \PMathML{}.
Other authors claim that precision can be lost by embracing content
because sometimes the actual layout used in a presentation or the name
used for variables are significant. \cite{GonzalesQUALIBETA} in
retrospect also described the choice of using \CMathML{} as a bad decision.
Other authors dismiss indexing of \CMathML{} because of the non-availability of libraries
or because of conversion from presentation to content being approximative and unreliable.
Finally, \cite{DBLP:conf/mkm/NghiemKTA14}
reports that automatic conversion of large formulae from Presentation to Content
may be computationally unfeasible,
and propose to limit the conversion to small ones.

Recently, the debate on presentation only vs parallel markup seems to be solved in favour of the latter.
For example, the system that scored better at the last NTCIR task
reports better scores when applied to content markup generated from \LaTeX{} w.r.t.
presentation only markup \cite{RuzickaMIAS}.
The authors second this observation already in \cite{LiskaSimilarity}.

Moreover, several works in the literature that deal with presentation markup enrich it
--- in the document itself or in the indexes ---
with additional annotations to make explicit additional semantics that is latent in the library \cite{DBLP:conf/mkm/Cairns04}
or in the text surrounding the formulae \cite{GonzalesQUALIBETA,KristiantoMCAT}. For example, in \cite{KristiantoMCAT}
artificial intelligence is applied to the whole document
to recover from the text surrounding the formulae the name associated to the mathematical entities in the formula
(e.g. ``posterior probability'', ``derivative of $f$''). \cite{GonzalesQUALIBETA} uses a cheaper approach by considering only one sentence around a formula, but it later observes that one sentence is often not sufficient and many relevant results are therefore missed.
Another example is an analysis of co-occurrence of symbols in the corpus to identify related ones.
It is shown that these techniques are important to augment recall or, sometimes, precision.
In our view, like the heuristic based presentation-to-content translation,
these are \emph{attempts to infer and store partial semantics of mathematical expressions}.
It may be questioned (see for example \cite{DBLP:conf/aisc/MillerY08})
if the current content markups (OpenMath and \CMathML{})
are the right instruments to augment presentation markup with partial, approximate semantics,
and if such additional semantics makes only sense in the indexes of search systems,
or it may be serialised to an XML format for being reused by third parties.

Systems based on \CMathML{}, parallel markup or semantics
appear in \cite{%
bancerek2006information,%
DBLP:conf/mkm/BancerekR03,%
DBLP:conf/mkm/BancerekU04,%
DBLP:conf/mkm/GauthierK14,%
HaginoPartial,%
HambasanMathwebsearch,%
KohlhaseMathwebsearch,%
1083259,%
LískaSojkaRuzickaMravec11,%
LiskaSimilarity,%
DBLP:conf/mkm/LiskaSR14,%
4530006,%
zbMATH05607875,%
munavalli2006mathfind,%
nguyen2012math,%
DBLP:conf/mkm/NormannK07,%
DBLP:conf/aisc/Rabe12,%
SchubotzQuerying,%
zbMATH05621231,%
DBLP:conf/mkm/ZhangY14%
}.

\section{Taxonomy of Techniques for Mathematical Retrieval}\label{sec:techiques}

Implementations of solutions to mathematical search problems can be obtained
combining one of the main techniques that will be presented in Section~\ref{sec:main}
with a choice of modular enhancement techniques from Section~\ref{sec:modular} used to improve precision, recall or both.

\subsection{Modular Enhancement Techniques}\label{sec:modular}

The following techniques are general enough to be applied to solve both the Document and the Formula Retrieval problems, and by analysing the literature it seems that every system eventually applies all of them.

\cscparagraph{Segmentation}\label{sec:segmentation}
A preliminary step to indexing is segmentation of documents into chunks.
Chunks are the unit of information to be returned to the user, with pointers to the parent document.
Segmentation is trivial on formal mathematical documents, hard on web-pages,
and intermediate on other resources like books or papers. See, for example, \cite{zhao2008math} for a discussion. Several systems implement segmentation; however, the last NTCIR competition has provided a data set of
already segmented documents \cite{aizawa2014ntcir} and that may hinder the study of segmentation techniques in the future.

\cscparagraph{Normalisation}\label{sec:normalisation}
To improve recall, both formulae in the query and the formulae in the
library are put in normal form before indexing them.
Having the same normal form is an equivalence relation $\equiv$,
and the query retrieves formulae up to $\equiv$.
For Formula Retrieval it is necessary that ${\equiv\,\R\,\equiv} \subseteq {\R}$.
For Document Retrieval the $\equiv$ relation must be compatible with the similarity and ranking functions.
When this is not the case, precision can be critically lowered.

Uses of normalisation include:
repairing of broken XML/MathML generated by automatic conversion tools \cite{DBLP:conf/mkm/MinerM07}
(e.g. when the structure imposed by \texttt{<mrows>} is not compatible with the mathematical structure);
removal of information that does not contribute to the semantics like comments, layout elements
(spaces, phantoms and linebreaks), XML/MathML attributes (color, font, elements in other
namespaces) \cite{FormanekEtAl:OpenMathUIWiP2012,4811843,DBLP:conf/mkm/MinerM07};
picking canonical representations of the same presentation/content when different MathML encodings are possible
(e.g. \texttt{msubsup} vs \texttt{msup} and \texttt{msub}, \texttt{mfenced}
vs use of two parentheses, applications of trigonometric functions with/without using parentheses,
etc.) \cite{altamimi2007more,FormanekEtAl:OpenMathUIWiP2012,DBLP:conf/mkm/MinerM07};
replacing names of bound variables with unique numerical indexes (e.g. De Brujin indexes)
to search up to $\alpha$-conversion \cite{DBLP:conf/mkm/MinerM07,DBLP:conf/mkm/NormannK07};
ignoring parentheses and ordering of arguments of associative/commutative
operators \cite{%
altamimi2007more,%
4530006,%
zbMATH05607875,%
DBLP:conf/mkm/NormannK07,%
DBLP:conf/icdim/ShatnawiY07,%
sojka2011art,%
youssef2006math%
};
expressing derived notions exposing the derivation
(e.g. replacing $x \geq y$ with $y \leq x$, $x \not\leq y$ with $\lnot (x \leq y)$, $\arcsin$ with
$\sin^{-1}$, etc.) \cite{altamimi2007more,DBLP:conf/mkm/MinerM07};
capturing logical equivalence/type isomorphisms (e.g. writing formulae in prenex normal form, currification of functions) \cite{delahaye2000information,DBLP:conf/mkm/GauthierK14,DBLP:conf/mkm/NormannK07}.

A normalised formula can be quite different from the original one, and that can be a symptom that the formula is not significant. Therefore in \cite{RuzickaMIAS} normalised formulae are weighted according to their similarity to the initial one, and weights are considered during the ranking phase with great results.

\cscparagraph{Approximation}
Normalisation does not lose information, converting a document to an equivalent one.
Many papers call ``normalisation'' an approximation phase where subformulae are replaced with constrained
placeholders to allow the formula to be matched by similarity.
For example: names of variables or constants can be replaced by a single name \cite{GaoICST};
all numeric constants by a single identifier \cite{GaoICST,DBLP:conf/mkm/MinerM07,sojka2011art};
subformulae may be replaced by their type.
For example, in \cite{HambasanMathwebsearch} type information is used to retrieve formulae by sorted unification,
i.e. by constraints with type placeholders in patterns. Approximation improves recall.
To limit the loss of precision, systems that approximate index both the original and the approximated formula
(or even several instances at different levels of approximation).
The effects of approximation are similar to those of query reduction,
but approximation is more efficient because it works at indexing time.

\cscparagraph{Enrichment}
Enrichment works on the library or on the query to augment the information stored/looked for in the index
by inferring new knowledge from existing one.
It can contribute to the solution of both the Document Retrieval and the Formula Retrieval problems. Typical examples of enrichment are:
heuristically generating and storing content metadata from \PMathML \cite{DBLP:conf/aisc/MillerY08};
automatic/interactive disambiguation of formulae in the queries
to perform a precise query at the content or semantics
level \cite{asperti2006content,bancerek2006information,DBLP:conf/mkm/BancerekR03,DBLP:conf/mkm/BancerekU04};
automatic inference of metadata from context analysis or usage analysis
(latent semantics) \cite{DBLP:conf/mkm/Cairns04,KristiantoMCAT,WolskaGrigore10}.

The most impressive application of enrichment is presented
in \cite{DBLP:conf/mkm/HaralambousQ14}.
The aim is to search for geometrical constructions that are described using a procedural language
(e.g. draw the segments connecting $A$ with $B$, $B$ with $C$, and $A$ with $C$).
Enrichment consists in replacing the procedural with a declarative description (e.g. $ABC$ is a triangle).
The same declarative description can be obtained by multiple procedural ones,
and thus recall is greatly improved.
The technique can also be seen as a form of normalisation (see Section~\ref{sec:normalisation})
where the normal form is not unique (e.g. it may be the case that by analysing the hypothesis
one could deduce that $ABC$ is also an equilateral triangle even if that is not stated in the
procedural description).

\cscparagraph{Query reduction}
Query reduction trades precision for recall by selectively dropping or weakening
some of the constraints present in the query.
Results obtained from reduced queries can be ranked after results from precise queries.
In the literature it occurs in many forms in solutions to both the Formula Retrieval and the Document Retrieval problems:
a constant can be weakened to other constants that co-occur frequently with the given one;
constants that occur too frequently can be dropped from the queries;
a formula may be required to match only the toplevel structure of the formula given as a query.

\subsection{Main Techniques}\label{sec:main}

The following techniques are mutually exclusive.
Moreover, each technique performs better on only one of the two problems.

\cscparagraph{Reduction to full-text searches}
The technology to perform full-text searches is very advanced
and there are popular open software implementations
with good performance like Apache Lucene/Solr and ElasticSearch.
The benefits of reducing search for formulae to full-text searches are speed of execution of the queries
and the combination of formula based and textual searches almost for free.
The main drawback is that the precise structure of a formula is partially lost
in the translations proposed in the literature,
and that it is impossible or very hard to capture precisely the kind of relations $\R$ used for
Formula Retrieval, unless $\R$ is approximated by a much coarser relation $\Rp$.
Therefore the technique has been successfully applied so far only to
Document Retrieval
\cite{%
adeel2008math,%
GaoICST,%
GonzalesQUALIBETA,%
4811843,%
KristiantoMCAT,%
libbrecht2006methods,%
LískaSojkaRuzickaMravec11,%
DBLP:conf/mkm/LiskaSR14,%
DBLP:conf/mkm/Miller13,%
DBLP:journals/amai/MillerY03,%
DBLP:conf/mkm/MinerM07,%
4530006,%
zbMATH05607875,%
PattaniyilCombining,%
munavalli2006mathfind,%
sojka2011art,%
DBLP:conf/iasse/Youssef05,%
DBLP:conf/mkm/Youssef07,%
DBLP:journals/mics/Youssef08%
}.\\
All the proposals employ vectors to represent features, and compare features with weighted cosine distance.
The usual approach consist in turning an expression into a (large) set of ``sentences'' that partially describe the
formula. For example, in \cite{KristiantoMCAT,TopicMCAT} a sentence is the set (ordered or not)
of symbols found in either a path from the root of the formula to a leaf, or as children of the same node.
Matching is then performed by a disjunctive query and results are ranked using TF-IDF and length normalisation.
As the authors claim, the system ``is \emph{too} flexible:
it is difficult to say where the relevant results stop and random matches begin;
thus we predict higher recall but lower precision rates than exact match systems''.
Other authors extract sentences or $n$-grams that capture the formula more precisely. As a general remark, the clear impression we got from the literature is that the fewer features extracted, the lower the precision.
All kinds of techniques can be used to extract the features, comprising regular expressions \cite{adeel2008math}
and finite state automata \cite{nguyen2012math}.

Some systems cluster documents at indexing time (e.g. \cite{adeel2008math}), and retrieve documents comparing
the feature vector of the query with the centroid of the cluster.
For example, documents about trigonometric functions are likely to be automatically clustered together.
However most systems do not seem to cluster in advance, and prefer the flexibility of weights to capture similarity
of features (e.g. similarity of occurrences of trigonometric functions).

According to the set of features extracted, the weighting function used, and the other modular techniques used in combination, the accuracy achieved by systems based on this technique range from extremely low to extremely high (see, for example,  \cite{aizawa2014ntcir}).

\cscparagraph{Structure-Based Indexing via Tries/Substitution Trees}\label{sec:trie}
Formula Retrieval can be solved with the data structures developed for automatic theorem proving to store libraries of
lemmas and quickly retrieve formulae up to instantiation/generalisation.
Pointers to all the statements are stored in the leaves of a tree that precisely encodes in its paths the
statements. To match a formula, the tree is recursively traversed using the formula
to drive the descent. The relations $\R$ that can be captured are only instantiation and
generalisation of whole formulae.
MathWebSearch \cite{HambasanMathwebsearch,KohlhaseMathwebsearch} and
\cite{Kamali:2010:NMR:1871437.1871635} are based on this approach.
Retrieval of formulae is very fast, assuming that the index can be entirely stored in main memory.

This approach consistently maximises precision but presents poor recall.
To accept larger relations, or to be applied to Document Retrieval, or to cope
with too rigid queries, the technique needs to be integrated with other ideas.
For example: to match subtrees of formulae in the library,
every subtree of a lemma needs to be stored as well in the index;
to solve unification problems up to an equational theory that admits normal forms,
all formulae are normalised;
to allow queries that use keywords or free text,
a free-text search engine must be run in parallel and the results need to be combined in the ranking
phase \cite{HambasanMathwebsearch,DBLP:conf/aisc/LibbrechtDMLDH08}.

\cscparagraph{Reduction to SQL or ad-hoc queries}
The third approach consists in approximating formulae via relations
to be stored in a relational DB \cite{DBLP:conf/mkm/AspertiS04,DBLP:conf/mkm/GuidiS03}.
An alternative consists in storing the relations in ad-hoc indexes in memory,
and it is employed when the indexes already exists for other purposes
(typically in libraries of formalised
knowledge) \cite{bancerek2006information,DBLP:conf/mkm/BancerekR03,DBLP:conf/mkm/BancerekU04}.
The technique is applied to Formula Retrieval and the database can be reused for Document Synthesis without modifications.
Approximated queries up to generalisation/instantiation can be made efficient \cite{DBLP:conf/mkm/AspertiS04}
without requiring an index stored in main memory for Structure-Based Indexing (see page~\pageref{sec:trie}).
Recall can be maximised by relaxing the representation of formulae as relations or by employing normalisation.
Ad-hoc inverted indices for paths and to map each \CMathML{} node to its parent have also been used
in \cite{HaginoPartial}: the search engine is very fast,
but the precision obtained is low.

\cscparagraph{Reduction to XML-based searches}
Some systems \cite{DBLP:conf/caine/AltamimiY07,DBLP:journals/mics/AltamimiY08,DBLP:conf/sede/YoussefA07}
that index MathML documents at the content level,
base their searching capabilities on the existing XPath/XQuery technology.
The system described in \cite{SchubotzQuerying}, which is based on Stratosphere, is batch oriented,
trades flexibility with performance, and it is essentially math-unaware
(for example, it does not normalise the input in any way).
Other systems \cite{DBLP:conf/mkm/CaprottiDT04,DBLP:conf/aisc/LibbrechtDMLDH08},
that deal with ontologies indexed in the Ontology Web Language,  
rely on third-party OWL search engines implementing graph matching.

\section{Ranking}\label{sec:ranking}
Because users only inspect the first results returned by a query, precision when solving Document Retrieval is strongly
determined by the ranking function. Ranking is also of paramount importance for Formula Retrieval:
when the search retrieves the candidates for progressing in a proof, correctly ranking the results may dramatically
cut the number of wrong proof attempts and backtracks. The ranking criterion for the two Problems is,
however, very different: for the first problem similarity of formulae in the query and in the results should
contribute significantly to the score; for the second problem the score should be determined by the intended use
of the results. For example, a lemma $L_1$ that exactly matches the goal to prove and has no premise should always
score better than a lemma $L_2$ that also exactly matches the goal, but that has hypotheses to be proved later.

Ranking according to the intended use for Formula Retrieval has received very little interest in the literature
we examined. On the other hand, several papers explicitly address ranking for Document Retrieval.
The consensus seem to be that a good ranking function needs to be sophisticated and that the usual metrics induced
by reduction to textual searches are completely inadequate (see, for example, \cite{DBLP:conf/mkm/Youssef07}).
All the proposed ranking techniques are strongly based on heuristics and, unfortunately, most of them are
incomparable and hard to combine. 

One class of metrics takes into consideration also the structure of
the formulae involved and the enriched semantics,
when available. For example, \cite{DBLP:conf/mkm/ZhangY14} heavily exploits \CMathML{} to rank results by
considering the taxonomic distance of constants (where close is approximated to being defined in the same content
dictionary), the data type hierarchical level (matching a function is more significant than matching a numerical
constant), matching depth (partially matching the formula at the top level is more significant than matching a
deeply nested subformula), coverage (percentage of formula matched), kind of matched expression (formula vs term).
All this information needs to be computed and amalgamated employing some kind of heuristic algorithm.
A second paper \cite{SchubotzEvaluation} confirmed that each one of the listed similarity feature factors significantly improves the ranking,
but the last one, that still contributes, has lower relevance.

In \cite{sojka2011art} ranking is determined by the weights used during matching, and the authors claim that each
document base and scientific field should have its own weighting function. Nevertheless, they
``tried to create a complex and robust weighting function that would be appropriate to many fields''.

In \cite{DBLP:conf/mkm/Youssef07} the author proposes a parameterised ranking function that works on mathematical
documents (not only formulae), that seems applicable to enriched presentation and that weights a lot of additional
information including keywords, the number of cross-references and their kind (e.g. definitional vs propositional).
Ranking employs a hybrid of scalarisation and vectorisation.

In \cite{DBLP:conf/mkm/KamaliT13} the authors propose to adopt tree edit distance to measure similarity of formulae.
Most of the paper is about optimisations to improve efficiency of ranking because tree edit distance is hard to
compute. The final proposal combines some clever memoisation and a procedure to quickly prune documents bounding
their similarity scores with a lightweight computation. The paper also shows benchmarks comparing the processing
time and success rate of most search and ranking algorithm in the literature, reimplemented by the authors and run
on the same dataset. From the benchmarks the method proposed seems to be superior, but the implementations do not
exploit relevant enriched information like cross-references, semantic proximity of definitions, etc.
The benchmarks are therefore non conclusive.

In \cite{nguyen2012math} the authors employ a continuous learning ranking model after having extracted features
from \CMathML{} mathematical formulae using a finite state automata. Benchmarks show their ranking to be
superior than the ones used in classical ranking of textual documents. However, they do not compare
with \cite{DBLP:conf/mkm/KamaliT13} or \cite{DBLP:conf/mkm/Youssef07} (that work on \PMathML).

Simpler approaches to ranking can be found in \cite{zbMATH05621231} (based on Subpath Set, reported to work well
only on ``simplified'' \CMathML{}) and \cite{zbMATH05621232} that works on \PMathML{} and measures
similarity as a function of the size of subtrees in common. The ranking metric used in \cite{adeel2008math} is a
TF-IDF modified with weights to assign more importance to some operators, but the details given to determine
the weights are insufficient.

Ranking algorithms can be too complex to be incorporated in the search phase, for example when using Lucene
technology. Moreover, they are typically slower than the search phase. Therefore several authors suggest to
re-rank only the first results of the query, that employs a simpler ranking measure to determine the interesting
candidates to be ranked more accurately \cite{DBLP:conf/mkm/KamaliT13,DBLP:conf/mkm/Youssef07}.

An algorithm to automatically categorise documents is presented in \cite{zhao2008math},
where it is argued that clustering documents according to their category greatly improves the usefulness of
the tool for the user.

\section{Evaluation of Math Information Retrieval}\label{sec:evaluation}
Several papers present benchmarks on the systems proposed, and rarely compare them with reimplementations of the
algorithms found in the literature (e.g. \cite{DBLP:conf/mkm/KamaliT13}).
The significance of most of these benchmarks is unclear, because conflicting results are found in the literature,
most techniques are not presented in sufficient details in the papers to be exactly reproduced,
and systems are very sensitive to the kind of queries examined.
The only alternative is to compare different tools on unbiased, standard benchmarks that are currently lacking.

The main issue is not to come up with large corpora of documents: at least for Document Retrieval on enriched
\PMathML{} documents, a large corpus can be easily obtained converting documents from ArXiV, DLMF,
PlanetMath, Wikipedia, etc. For Formula Retrieval, the existing libraries of interactive theorem provers,
like Mizar and Coq, can be directly used after conversion.
The problem is to determine large sets of \emph{real world, interesting queries}, and to evaluate the results.
Automatic evaluation is particularly hard in the domain of mathematics, whereas manual evaluation is limited to a
tiny number of queries and runs. Formulating good sets of queries is also complex, because users with different
mathematical background and motivations are likely to issue different queries. Moreover, what makes a query hard
can just be the use of non standard mathematical notations, errors in the encoding of formulae,
or formulation at the wrong level of abstraction. \cite{1083259} discusses the problem at length and reviews
the state of the art of evaluation of Math Information Retrieval before 2013, including the experience of the
MIR workshop at CICM 2012 were two systems were compared on about ten hard queries proposed by the judges,
and the conclusion was that the systems were too sensitive to the formulation of the query.

The situation is improving since 2013 with the creation of a math oriented task in the NTCIR
initiative \cite{aizawa2013ntcir,aizawa2014ntcir}
that is attracting a small, but increasing number of participants
\cite{%
GaoICST,%
GonzalesQUALIBETA,%
HaginoPartial,%
HambasanMathwebsearch,%
KohlhaseMathwebsearch,%
KristiantoMCAT,%
LarsonBerkeley,%
LipaniTUW,%
LiskaSimilarity,%
PattaniyilCombining,%
RuzickaMIAS,%
SchubotzQuerying,%
SchubotzEvaluation,%
TopicMCAT%
}.
The initiative is too young to come to definite conclusions and the current choice of tasks and queries is not
granted yet to have significant coverage and to be unbiased. For example,
in \cite{KristiantoMCAT} the authors report that despite several improvements
to the tool (quantified via NTCIR-11 runs), their tool scored lower than in NTCIR-10.
They justify the phenomenon by noticing that ``in NTCIR-11, query variables get much bigger emphasis,
most topics feature complete and very particular formulae, and sub-formulae matching is not nearly as useful
as before''. Indeed, as reported in \cite{aizawa2014ntcir} ``the design decision \ldots
to exclusively concentrate on formula/keyword queries and use paragraphs as retrieval units \ldots
has also focused research away from questions like result presentation and user interaction. \ldots
few of the systems has invested into further semantics extraction from the data set. \ldots
We feel that this direction should be addressed more in future challenges''.
An effect of the bias towards search up to unification w.r.t. search up to similarity is observable
in \cite{PattaniyilCombining} too: the system proposed works very well
even if it works on \PMathML{} only and the set of features extracted is very simple (bag-of-symbol-pairs model, where a pair is made of two symbols in a father-son relation). The reason why it works well is that the authors also index approximated pairs where the child is a wildcard, and in the future works they are thinking at improving even more the handling of wildcards to score better. In comparison, most other systems based on feature vectors just replace wildcards in the query with \texttt{<m:ci>} identifiers.
The emphasis on formula queries is also to be evaluated considering the already cited works that conclude that users do not see (yet?) much value in them \cite{DBLP:conf/mkm/KohlhaseK07,zhao2008math}.

Finally, the NTCIR task does not cover distinctly the Document Retrieval and the Formula Retrieval problems, but only Document Retrieval with an emphasis on exact pattern matching of formulae that should be more distinctive of Formula Retrieval.

\section{Availability of Math Retrieval Systems}
Most of the systems described in the literature are research
prototypes, and the majority of them are no longer working or no
longer accessible. At the time this paper was written, the only ones
for Document Retrieval with a running Web interface or code that can be downloaded are: 1) Design Science's MathDex (formerly MathFind) 2) NIST DMLF 3) MathWebSearch 4) MiAS (Math Indexer and Searcher), also used to search the EuDML 5) the system described in \cite{PattaniyilCombining}. In addition to those, the following commercial systems are also accessible: a) Springer LaTeXSearch b) Wolfram Alpha. Systems a), 1), 2) and 5) are based on \PMathML{}; systems 3) and 4) can use either \PMathML{}
or \CMathML/parallel markup, but work better on the latter; finally system b) actually generates on the fly most of the result of the query, for example by plotting functions, computing their Taylor expansion, etc. It does not really qualify then as a search engine.

Most interactive theorem provers also have their own implementation of a search engine to solve Formula Retrieval. Most of the time, the implementation is embedded in the system and does not work on the whole library at once, with the exception of MML Query for Mizar.

\section{Conclusions}
Mathematical knowledge retrieval, the low hanging fruit of Mathematical Knowledge Management, is still far from being grasped. Despite the significant amount of work dedicated to the topic in the last 12 years, only a few systems are still available, and their precision and recall scores compared to other knowledge retrieval fields are low. Moreover, usability and user requirement studies suggest that queries containing formulae --- the main focus of the majority of papers --- are perceived by users as not very useful (yet?).

The main contributions of this paper have been providing an hopefully comprehensive bibliography on the subject, and presenting taxonomies for both mathematical retrieval problems and techniques. We believe that our purpose driven taxonomy can be useful in classifying papers, in clarifying the scope of application of techniques and in the much needed development of unbiased benchmarks for mathematical retrieval.

\bibliographystyle{alpha} 
\bibliography{references1,references2,references3,references4,references5}

\newcommand{\etalchar}[1]{$^{#1}$}
\begin{thebibliography}{AGSC{\etalchar{+}}06}

\bibitem[ACK08]{adeel2008math}
Muhammad Adeel, Hui~Siu Cheung, and Sikandar~Hayat Khiyal.
\newblock {Math GO! prototype of a content based mathematical formula search
  engine}.
\newblock {\em Journal of Theoretical and Applied Information Technology},
  4(10):1002--1012, 2008.

\bibitem[AGSC{\etalchar{+}}06]{asperti2006content}
Andrea Asperti, Ferruccio Guidi, Claudio Sacerdoti~Coen, Enrico Tassi, and
  Stefano Zacchiroli.
\newblock {A content based mathematical search engine: Whelp}.
\newblock In {\em Types for Proofs and Programs}, pages 17--32. Springer, 2006.

\bibitem[AKO13]{aizawa2013ntcir}
Akiko Aizawa, Michael Kohlhase, and Iadh Ounis.
\newblock {NTCIR-10 math pilot task overview}.
\newblock In {\em Proceedings of the 10th NTCIR Conference, Tokyo, Japan},
  pages 654--661, 2013.

\bibitem[AKO14]{aizawa2014ntcir}
Akiko Aizawa, Michael Kohlhase, and Iadh Ounis.
\newblock {NTCIR-11 Math 2 Task Overview}.
\newblock In {\em Proc. 10th NTCIR Conference (Tokyo, Japan)}, pages 88--98,
  2014.

\bibitem[AS04]{DBLP:conf/mkm/AspertiS04}
Andrea Asperti and Matteo Selmi.
\newblock {Efficient Retrieval of Mathematical Statements}.
\newblock In {\em Mathematical Knowledge Management, Third International
  Conference, MKM 2004, Bialowieza, Poland, September 19-21, 2004,
  Proceedings}, pages 17--31, 2004.

\bibitem[AY07a]{altamimi2007more}
Moody~E. Altamimi and Abdou Youssef.
\newblock {A more canonical form of content MathML to facilitate math search}.
\newblock In {\em Proc. Extreme Markup Languages}, 2007.

\bibitem[AY07b]{DBLP:conf/caine/AltamimiY07}
Moody~Ebrahem Altamimi and Abdou~S. Youssef.
\newblock {Wildcards in Math Search, Implementation Issues}.
\newblock In {\em Proceedings of the ISCA 20th International Conference on
  Computer Applications in Industry and Engineering, CAINE 2007, November 7-9,
  2007, San Francisco, California, USA}, pages 90--96, 2007.

\bibitem[AY08a]{zbMATH05607874}
Seyed~Ali {Ahmadi} and Abdou {Youssef}.
\newblock {Lexical error compensation in handwritten-based mathematical
  information retrieval.}
\newblock In {\em {DML 2008. Towards digital mathematics library, Birmingham,
  UK, July 27th, 2008. Proceedings}}, pages 43--54. Brno: Masaryk University,
  2008.

\bibitem[AY08b]{DBLP:journals/mics/AltamimiY08}
Moody~Ebrahem Altamimi and Abdou~S. Youssef.
\newblock {A Math Query Language with an Expanded Set of Wildcards}.
\newblock {\em Mathematics in Computer Science}, 2(2):305--331, 2008.

\bibitem[AZ04]{Asperti04searchingmathematics}
Andrea Asperti and Stefano Zacchiroli.
\newblock {Searching Mathematics on the Web: State of the Art and Future
  Developments}.
\newblock In FIZ, editor, {\em Proceedings of New Developments in Electronic
  Publishing of Mathematics}, pages 9--18, 2004.

\bibitem[Ban06]{bancerek2006information}
Grzegorz Bancerek.
\newblock {Information retrieval and rendering with MML Query}.
\newblock In {\em Mathematical Knowledge Management}, pages 266--279. Springer,
  2006.

\bibitem[BF03]{baumgartner2003automated}
Peter Baumgartner and Ulrich Furbach.
\newblock {Automated deduction techniques for the management of personalized
  documents}.
\newblock {\em Annals of Mathematics and Artificial Intelligence},
  38(1-3):211--228, 2003.

\bibitem[BR03]{DBLP:conf/mkm/BancerekR03}
Grzegorz Bancerek and Piotr Rudnicki.
\newblock {Information Retrieval in MML}.
\newblock In {\em Mathematical Knowledge Management, Second International
  Conference, MKM 2003, Bertinoro, Italy, February 16-18, 2003, Proceedings},
  pages 119--132, 2003.

\bibitem[BU04]{DBLP:conf/mkm/BancerekU04}
Grzegorz Bancerek and Josef Urban.
\newblock {Integrated Semantic Browsing of the Mizar Mathematical Library for
  Authoring Mizar Articles}.
\newblock In {\em Mathematical Knowledge Management, Third International
  Conference, MKM 2004, Bialowieza, Poland, September 19-21, 2004,
  Proceedings}, pages 44--57, 2004.

\bibitem[Cai04]{DBLP:conf/mkm/Cairns04}
Paul~A. Cairns.
\newblock {Informalising Formal Mathematics: Searching the Mizar Library with
  Latent Semantics}.
\newblock In {\em Mathematical Knowledge Management, Third International
  Conference, MKM 2004, Bialowieza, Poland, September 19-21, 2004,
  Proceedings}, pages 58--72, 2004.

\bibitem[CDT04]{DBLP:conf/mkm/CaprottiDT04}
Olga Caprotti, Mike Dewar, and Daniele Turi.
\newblock {Mathematical Service Matching Using Description Logic and OWL}.
\newblock In {\em Mathematical Knowledge Management, Third International
  Conference, MKM 2004, Bialowieza, Poland, September 19-21, 2004,
  Proceedings}, pages 73--87, 2004.

\bibitem[Del00]{delahaye2000information}
David Delahaye.
\newblock {Information retrieval in a coq proof library using type
  isomorphisms}.
\newblock In {\em Types for Proofs and Programs}, pages 131--147. Springer,
  2000.

\bibitem[FLRS12]{FormanekEtAl:OpenMathUIWiP2012}
David Form\'anek, Martin L\'{\i}\v{s}ka, Michal R\r{u}\v{z}i\v{c}ka, and Petr
  Sojka.
\newblock {Normalization of Digital Mathematics Library Content}.
\newblock In James Davenport, Johan Jeuring, Christoph Lange, and Paul
  Libbrecht, editors, {\em Joint Proceedings of the 24th OpenMath Workshop, the
  7th Workshop on Mathematical User Interfaces (MathUI), and the Work in
  Progress Section of the Conference on Intelligent Computer Mathematics},
  number 921 in CEUR Workshop Proceedings, pages 91--103, Aachen, 2012.

\bibitem[GK14]{DBLP:conf/mkm/GauthierK14}
Thibault Gauthier and Cezary Kaliszyk.
\newblock {Matching Concepts across HOL Libraries}.
\newblock In {\em Intelligent Computer Mathematics - International Conference,
  CICM 2014, Coimbra, Portugal, July 7-11, 2014. Proceedings}, pages 267--281,
  2014.

\bibitem[GPBB14]{GonzalesQUALIBETA}
Jos\'e~Mar\'{\i}a Gonz\'alez~Pinto, Simon Barthel, and Wolf-Tilo Balke.
\newblock {QUALIBETA at the NTCIR-11 Math 2 Task: An Attempt to Query Math
  Collections}.
\newblock In {\em Proc. 10th NTCIR Conference (Tokyo, Japan)}, pages 103--107,
  2014.

\bibitem[GS03]{DBLP:conf/mkm/GuidiS03}
Ferruccio Guidi and Irene Schena.
\newblock {A Query Language for a Metadata Framework about Mathematical
  Resources}.
\newblock In {\em Mathematical Knowledge Management, Second International
  Conference, MKM 2003, Bertinoro, Italy, February 16-18, 2003, Proceedings},
  pages 105--118, 2003.

\bibitem[GWHT14]{GaoICST}
Liangcai Gao, Yuehan Wang, Leipeng Hao, and Zhi Tang.
\newblock {ICST Math Retrieval System for NTCIR-11 Math-2 Task}.
\newblock In {\em Proc. 10th NTCIR Conference (Tokyo, Japan)}, pages 99--102,
  2014.

\bibitem[HHN08]{4811843}
H.~Hashimoto, Y.~Hijikata, and S.~Nishida.
\newblock {Incorporating breadth first search for indexing MathML objects}.
\newblock In {\em Systems, Man and Cybernetics, 2008. SMC 2008. IEEE
  International Conference on}, pages 3519--3523, Oct 2008.

\bibitem[HKP14]{HambasanMathwebsearch}
Radu Hambasan, Michael Kohlhase, and Corneliu Prodescu.
\newblock {MathWebSearch at NTCIR-11}.
\newblock In {\em Proc. 10th NTCIR Conference (Tokyo, Japan)}, pages 114--119,
  2014.

\bibitem[HQ14]{DBLP:conf/mkm/HaralambousQ14}
Yannis Haralambous and Pedro Quaresma.
\newblock {Querying Geometric Figures Using a Controlled Language, Ontological
  Graphs and Dependency Lattices}.
\newblock In {\em Intelligent Computer Mathematics - International Conference,
  CICM 2014, Coimbra, Portugal, July 7-11, 2014. Proceedings}, pages 298--311,
  2014.

\bibitem[HS13]{HaginoPartial}
Hiroya Hagino and Hiroaki Saito.
\newblock {Partial-match Retrieval with Structure-reflected Indices at the
  NTCIR-10 Math Task}.
\newblock In {\em Proc. 10th NTCIR Conference (Tokyo, Japan)}, pages 692--695,
  2013.

\bibitem[KK07]{DBLP:conf/mkm/KohlhaseK07}
Andrea Kohlhase and Michael Kohlhase.
\newblock {\emph{Re} examining the MKM Value Proposition: From Math Web Search
  to Math Web \emph{Re} Search}.
\newblock In {\em Towards Mechanized Mathematical Assistants, 14th Symposium,
  Calculemus 2007, 6th International Conference, MKM 2007, Hagenberg, Austria,
  June 27-30, 2007, Proceedings}, pages 313--326, 2007.

\bibitem[Koh14]{DBLP:conf/mkm/Kohlhase14}
Andrea Kohlhase.
\newblock {Search Interfaces for Mathematicians}.
\newblock In {\em Intelligent Computer Mathematics - International Conference,
  CICM 2014, Coimbra, Portugal, July 7-11, 2014. Proceedings}, pages 153--168,
  2014.

\bibitem[KP13]{KohlhaseMathwebsearch}
Michael Kohlhase and Corneliu Prodescu.
\newblock {MathWebSearch at NTCIR-10}.
\newblock In {\em Proc. 10th NTCIR Conference (Tokyo, Japan)}, pages 675--679,
  2013.

\bibitem[KT09]{zbMATH05621232}
Shahab {Kamali} and Frank~Wm. {Tompa}.
\newblock {Improving mathematics retrieval.}
\newblock In {\em {DML 2009. Towards digital mathematics library, Grand Bend,
  Ontario, Canada, July 8--9th 2009. Proceedings}}, pages 37--48. Brno: Masaryk
  University, 2009.

\bibitem[KT10]{Kamali:2010:NMR:1871437.1871635}
Shahab Kamali and Frank~Wm. Tompa.
\newblock {A New Mathematics Retrieval System}.
\newblock In {\em Proceedings of the 19th ACM International Conference on
  Information and Knowledge Management}, CIKM `10, pages 1413--1416, New York,
  NY, USA, 2010. ACM.

\bibitem[KT13]{DBLP:conf/mkm/KamaliT13}
Shahab Kamali and Frank~Wm. Tompa.
\newblock {Structural Similarity Search for Mathematics Retrieval}.
\newblock In {\em Intelligent Computer Mathematics - MKM, Calculemus, DML, and
  Systems and Projects 2013, Held as Part of CICM 2013, Bath, UK, July 8-12,
  2013. Proceedings}, pages 246--262, 2013.

\bibitem[KTHA14]{KristiantoMCAT}
Giovanni~Yoko Kristianto, Goran Topi\'c, Florence Ho, and Akiko Aizawa.
\newblock {The MCAT math retrieval system for NTCIR-11 math track}.
\newblock In {\em Proc. 11th NTCIR Conference (Tokyo, Japan)}, pages 120--126,
  2014.

\bibitem[L\'10]{bachliska}
Martin L\'{\i}\v{s}ka.
\newblock {Searching Mathematical Texts}, 2010.

\bibitem[L\'13]{1083259}
Martin L\'{\i}\v{s}ka.
\newblock {Evaluation of Mathematics Retrieval}, 2013.

\bibitem[LAP{\etalchar{+}}14]{LipaniTUW}
Aldo Lipani, Linda Andersson, Florina Piroi, Mihai Lupu, and Allan Hanbury.
\newblock {TUW-IMP at the NTCIR-11 Math-2}.
\newblock In {\em Proc. 11th NTCIR Conference (Tokyo, Japan)}, pages 143--146,
  2014.

\bibitem[LDM{\etalchar{+}}08]{DBLP:conf/aisc/LibbrechtDMLDH08}
Paul Libbrecht, Cyrille Desmoulins, Christian Mercat, Colette Laborde, Michael
  Dietrich, and Maxim Hendriks.
\newblock {Cross-Curriculum Search for Intergeo}.
\newblock In {\em Intelligent Computer Mathematics, 9th International
  Conference, AISC 2008, 15th Symposium, Calculemus 2008, 7th International
  Conference, MKM 2008, Birmingham, UK, July 28 - August 1, 2008. Proceedings},
  pages 520--535, 2008.

\bibitem[Lib13]{DBLP:conf/mkm/Libbrecht13}
Paul Libbrecht.
\newblock {Escaping the Trap of Too Precise Topic Queries}.
\newblock In {\em Intelligent Computer Mathematics - MKM, Calculemus, DML, and
  Systems and Projects 2013, Held as Part of CICM 2013, Bath, UK, July 8-12,
  2013. Proceedings}, pages 296--309, 2013.

\bibitem[LM06]{libbrecht2006methods}
Paul Libbrecht and Erica Melis.
\newblock Methods to access and retrieve mathematical content in activemath.
\newblock In {\em Mathematical Software-ICMS 2006}, pages 331--342. Springer,
  2006.

\bibitem[LRG13]{LarsonBerkeley}
Ray~R. Larson, Chloe~J. Reynolds, and Fredric~C. Gey.
\newblock {The Abject Failure of Keyword IR for Mathematics Search: Berkeley at
  NTCIR-10 Math}.
\newblock In {\em Proc. 10th NTCIR Conference (Tokyo, Japan)}, pages 662--666,
  2013.

\bibitem[LSLM11]{LískaSojkaRuzickaMravec11}
Martin {L\'{\i}\v{s}ka}, Petr {Sojka}, Michal {L\'{\i}\v{s}ka}, and Petr
  {Mravec}.
\newblock {Web Interface and Collection for Mathematical Retrieval: WebMIaS and
  MREC.}
\newblock In {\em {DML 2011. Towards digital mathematics library, Bertinoro,
  Italy, July 20-21st, 2011. Proceedings}}, pages 77--84. Brno: Masaryk
  University, 2011.

\bibitem[LSR13]{LiskaSimilarity}
Martin L\'{\i}\v{s}ka, Petr Sojka, and Michal R\r{u}\v{z}i\v{c}ka.
\newblock {Similarity Search for Mathematics: Masaryk University team at the
  NTCIR-10 Math Task}.
\newblock In {\em Proc. 10th NTCIR Conference (Tokyo, Japan)}, pages 686--691,
  2013.

\bibitem[LSR14]{DBLP:conf/mkm/LiskaSR14}
Martin L\'{\i}\v{s}ka, Petr Sojka, and Michal R\r{u}\v{z}i\v{c}ka.
\newblock {Math Indexer and Searcher Web Interface - Towards Fulfillment of
  Mathematicians' Information Needs}.
\newblock In {\em Intelligent Computer Mathematics - International Conference,
  CICM 2014, Coimbra, Portugal, July 7-11, 2014. Proceedings}, pages 444--448,
  2014.

\bibitem[MG08a]{4530006}
J.~Misutka and L.~Galambos.
\newblock {Mathematical Extension of Full Text Search Engine Indexer}.
\newblock In {\em Information and Communication Technologies: From Theory to
  Applications, 2008. ICTTA 2008. 3rd International Conference on}, pages 1--6,
  April 2008.

\bibitem[MG08b]{zbMATH05607875}
Jozef {Mi\v{s}utka} and Leo {Galambo\v{s}}.
\newblock {Extending full text search engine for mathematical content.}
\newblock In {\em {DML 2008. Towards digital mathematics library, Birmingham,
  UK, July 27th, 2008. Proceedings}}, pages 55--67. Brno: Masaryk University,
  2008.

\bibitem[Mil13]{DBLP:conf/mkm/Miller13}
Bruce~R. Miller.
\newblock {Three Years of DLMF: Web, Math and Search}.
\newblock In {\em Intelligent Computer Mathematics - MKM, Calculemus, DML, and
  Systems and Projects 2013, Held as Part of CICM 2013, Bath, UK, July 8-12,
  2013. Proceedings}, pages 288--295, 2013.

\bibitem[MM06]{munavalli2006mathfind}
Rajesh Munavalli and Robert Miner.
\newblock {Mathfind: a math-aware search engine}.
\newblock In {\em Proceedings of the 29th annual international ACM SIGIR
  conference on Research and development in information retrieval}, pages
  735--735. ACM, 2006.

\bibitem[MM07]{DBLP:conf/mkm/MinerM07}
Robert Miner and Rajesh Munavalli.
\newblock {An Approach to Mathematical Search Through Query Formulation and
  Data Normalization}.
\newblock In {\em Towards Mechanized Mathematical Assistants, 14th Symposium,
  Calculemus 2007, 6th International Conference, MKM 2007, Hagenberg, Austria,
  June 27-30, 2007, Proceedings}, pages 342--355, 2007.

\bibitem[MY03]{DBLP:journals/amai/MillerY03}
Bruce~R. Miller and Abdou Youssef.
\newblock {Technical Aspects of the Digital Library of Mathematical Functions}.
\newblock {\em Ann. Math. Artif. Intell.}, 38(1-3):121--136, 2003.

\bibitem[MY08]{DBLP:conf/aisc/MillerY08}
Bruce~R. Miller and Abdou Youssef.
\newblock {Augmenting Presentation MathML for Search}.
\newblock In {\em Intelligent Computer Mathematics, 9th International
  Conference, {AISC} 2008, 15th Symposium, Calculemus 2008, 7th International
  Conference, MKM 2008, Birmingham, UK, July 28 - August 1, 2008. Proceedings},
  pages 536--542, 2008.

\bibitem[NCH12]{nguyen2012math}
Tam~T Nguyen, Kuiyu Chang, and Siu~Cheung Hui.
\newblock {A math-aware search engine for math question answering system}.
\newblock In {\em Proceedings of the 21st ACM international conference on
  Information and knowledge management}, pages 724--733. ACM, 2012.

\bibitem[NK07]{DBLP:conf/mkm/NormannK07}
Immanuel Normann and Michael Kohlhase.
\newblock {Extended Formula Normalization for \emph{epsilon} -Retrieval and
  Sharing of Mathematical Knowledge}.
\newblock In {\em Towards Mechanized Mathematical Assistants, 14th Symposium,
  Calculemus 2007, 6th International Conference, MKM 2007, Hagenberg, Austria,
  June 27-30, 2007, Proceedings}, pages 356--370, 2007.

\bibitem[NKTA14]{DBLP:conf/mkm/NghiemKTA14}
Minh-Quoc Nghiem, Giovanni~Yoko Kristianto, Goran Topi\'c, and Akiko Aizawa.
\newblock {Which One Is Better: Presentation-Based or Content-Based Math
  Search?}
\newblock In {\em Intelligent Computer Mathematics - International Conference,
  CICM 2014, Coimbra, Portugal, July 7-11, 2014. Proceedings}, pages 200--212,
  2014.

\bibitem[PZ14]{PattaniyilCombining}
Nidhin Pattaniyil and Richard Zanibbi.
\newblock {Combining TF-IDF Text Retrieval with an Inverted Index over Symbol
  Pairs in Math Expressions: The Tangent Math Search Engine at NTCIR 2014}.
\newblock In {\em Proc. 11th NTCIR Conference (Tokyo, Japan)}, pages 135--142,
  2014.

\bibitem[Rab12]{DBLP:conf/aisc/Rabe12}
Florian Rabe.
\newblock {A Query Language for Formal Mathematical Libraries}.
\newblock In {\em Intelligent Computer Mathematics - 11th International
  Conference, AISC 2012, 19th Symposium, Calculemus 2012, 5th International
  Workshop, DML 2012, 11th International Conference, MKM 2012, Systems and
  Projects, Held as Part of CICM 2012, Bremen, Germany, July 8-13, 2012.
  Proceedings}, pages 143--158, 2012.

\bibitem[RSL14]{RuzickaMIAS}
Michal R\r{u}\v{z}i\v{c}ka, Petr Sojka, and Martin L\'{\i}\v{s}ka.
\newblock {Math Indexer and Searcher under the Hood: History and Development of
  a Winning Strategy}.
\newblock In {\em Proc. 11th NTCIR Conference (Tokyo, Japan)}, pages 127--134,
  2014.

\bibitem[SL11]{sojka2011art}
Petr Sojka and Martin L\'{\i}\v{s}ka.
\newblock {The art of mathematics retrieval}.
\newblock In {\em Proceedings of the 11th ACM symposium on Document
  engineering}, pages 57--60. ACM, 2011.

\bibitem[SLM13]{SchubotzQuerying}
Moritz Schubotz, Marcus Leich, and Volker Markl.
\newblock {Querying large Collections of Mathematical Publications: NTCIR10
  Math Task}.
\newblock In {\em Proc. 10th NTCIR Conference (Tokyo, Japan)}, pages 667--674,
  2013.

\bibitem[SY07]{DBLP:conf/icdim/ShatnawiY07}
Mohammed Shatnawi and Abdou Youssef.
\newblock {Equivalence detection using parse-tree normalization for math
  search}.
\newblock In {\em Second IEEE International Conference on Digital Information
  Management (ICDIM), December 11-13, 2007, Lyon, France, Proceedings}, pages
  643--648, 2007.

\bibitem[SYM{\etalchar{+}}14]{SchubotzEvaluation}
Moritz Schubotz, Abdou Youssef, Volker Markl, Howard~S. Cohl, and Jimmy~J. Li.
\newblock {Evaluation of Similarity-Measure Factors for Formulae based on the
  NTCIR-11 Math Task}.
\newblock In {\em Proc. 10th NTCIR Conference (Tokyo, Japan)}, pages 108--113,
  2014.

\bibitem[TKNA13]{TopicMCAT}
Goran Topi\'c, Giovanni~Yoko Kristianto, Minh-Quoc Nghiem, and Akiko Aizawa.
\newblock {The MCAT math retrieval system for NTCIR-10 math track}.
\newblock In {\em Proc. 10th NTCIR Conference (Tokyo, Japan)}, pages 680--685,
  2013.

\bibitem[WG10]{WolskaGrigore10}
Magdalena {Wolska} and Mihai {Grigore}.
\newblock {Symbol Declarations in Mathematical Writing.}
\newblock In {\em {DML 2010. Towards digital mathematics library, Paris,
  France, July 7-8th, 2010. Proceedings}}, pages 119--127. Brno: Masaryk
  University, 2010.

\bibitem[YA07]{DBLP:conf/sede/YoussefA07}
Abdou~S. Youssef and Moody~Ebrahem Altamimi.
\newblock {An extensive math query language}.
\newblock In {\em 16th International Conference on Software Engineering and
  Data Engineering (SEDE-2007), July 9-11, 2007, Imperial Palace Hotel Las
  Vegas, Las Vegas, Nevada, USA, Proceedings}, pages 57--63, 2007.

\bibitem[YA09]{zbMATH05621231}
Keisuke {Yokoi} and Akiko {Aizawa}.
\newblock {An approach to similarity search for mathematical expressions using
  MathML.}
\newblock In {\em {DML 2009. Towards digital mathematics library, Grand Bend,
  Ontario, Canada, July 8--9th 2009. Proceedings}}, pages 27--35. Brno: Masaryk
  University, 2009.

\bibitem[You05]{DBLP:conf/iasse/Youssef05}
Abdou Youssef.
\newblock {Search of Mathematical Contents: Issues And Methods}.
\newblock In {\em Proceedings of the ISCA 14th International Conference on
  Intelligent and Adaptive Systems and Software Engineering, July 20-22, 2005,
  Novotel Toronto Centre, Toronto, Canada}, pages 100--105, 2005.

\bibitem[You06]{DBLP:conf/mkm/Youssef06}
Abdou Youssef.
\newblock {Roles of Math Search in Mathematics}.
\newblock In {\em Mathematical Knowledge Management, 5th International
  Conference, {MKM} 2006, Wokingham, UK, August 11-12, 2006, Proceedings},
  pages 2--16, 2006.

\bibitem[You07]{DBLP:conf/mkm/Youssef07}
Abdou Youssef.
\newblock {Methods of Relevance Ranking and Hit-content Generation in Math
  Search}.
\newblock In {\em Towards Mechanized Mathematical Assistants, 14th Symposium,
  Calculemus 2007, 6th International Conference, MKM 2007, Hagenberg, Austria,
  June 27-30, 2007, Proceedings}, pages 393--406, 2007.

\bibitem[You08]{DBLP:journals/mics/Youssef08}
Abdou~S. Youssef.
\newblock {Relevance Ranking and Hit Description in Math Search}.
\newblock {\em Mathematics in Computer Science}, 2(2):333--353, 2008.

\bibitem[YS06]{youssef2006math}
Abdou Youssef and Mohammed Shatnawi.
\newblock {Math search with equivalence detection using parse-tree
  normalization}.
\newblock In {\em The 4th International Conference on Computer Science and
  Information Technology}, 2006.

\bibitem[ZB12]{zanibbi2012recognition}
Richard Zanibbi and Dorothea Blostein.
\newblock {Recognition and retrieval of mathematical expressions}.
\newblock {\em International Journal on Document Analysis and Recognition
  (IJDAR)}, 15(4):331--357, 2012.

\bibitem[ZKT08]{zhao2008math}
Jin Zhao, Min-Yen Kan, and Yin~Leng Theng.
\newblock {Math information retrieval: user requirements and prototype
  implementation}.
\newblock In {\em Proceedings of the 8th ACM/IEEE-CS joint conference on
  Digital libraries}, pages 187--196. ACM, 2008.

\bibitem[ZY14]{DBLP:conf/mkm/ZhangY14}
Qun Zhang and Abdou Youssef.
\newblock {An Approach to Math-Similarity Search}.
\newblock In {\em Intelligent Computer Mathematics - International Conference,
  {CICM} 2014, Coimbra, Portugal, July 7-11, 2014. Proceedings}, pages
  404--418, 2014.

\end{thebibliography}

\end{document}